\documentclass[a4paper,11pt]{article}
    \usepackage{cite}
    \usepackage{amsmath,amssymb}
    \usepackage{bm}
    \usepackage{mathrsfs}
    \usepackage[dvipdfm]{graphicx}
    \usepackage[abs]{overpic}
    \usepackage{pict2e}
    \newcommand{\be}{\begin{equation}}
    \newcommand{\ee}{\end{equation}}
    \newcommand{\nbe}{\begin{equation*}}
    \newcommand{\nee}{\end{equation*}}
    \newcommand{\gravcoup}{\kappa_4^2}

    \newcommand{\pd}{\partial}
    
    \newcommand{\half}{\frac{1}{2}}
    \usepackage{layout}

\begin{document}
\title{\textbf{Mutated hybrid inflation in $f(R,\Box R)$-gravity} \\ \phantom{f(R,boxR)}}
\author{\Large{Masao Iihoshi}\footnote{Electronic address: iihoshi@kiso.phys.se.tmu.ac.jp}
                \\ \\
                \emph{Department of Physics, Tokyo Metropolitan University,} \\
    \emph{Hachioji, Tokyo 192-0397, Japan}}
\date{\empty} 
\maketitle

\begin{abstract}
A new hybrid inflationary scenario in the context of $f(R,\Box R)$-gravity is proposed.
Demanding the waterfall field to `support the potential from below' [unlike the original proposal by Stewart in Phys. Lett. \textbf{B345}, 414 (1995)],
we demonstrate that the scalar potential is similar to that of the large-field chaotic inflation model proposed by Linde in Phys. Lett. \textbf{B129}, 177 (1983).
Inflationary observables are used to constrain the parameter space of our model;
in the process, an interesting limit on the number of $e$-folds $N$ is found.
\end{abstract}

\section{Introduction}
Nowadays various modifications of General Relativity (GR) are actively studied, since it may explain the rapid accelerated expansion in the very early stage of the Universe
(i.e. inflation) and the late time acceleration, without introducing matter fields of desired properties.
Inflation is now the standard paradigm for cosmology, since it elegantly resolves serious philosophical problems\cite{Guth}
and can explain the present density perturbation of the Universe\cite{Komatsu.etal}.
It seems to be quite natural to try to describe inflationary epoch in fundamental physics in which the mechanism of inflation naturally exists.

One such candidate is supergravity (SUGRA), in which many scalar fields (would-be inflaton) arise as the superpartners of fermionic fields.
SUGRA is known to be the low energy effective theory of superstring/M-theory.\footnote{There are many models of inflation derived from superstring/M-theory,
see, e.g. Ref.\cite{Erd}.}
It looks natural to use SUGRA to describe inflation\cite{MSYY94,KL10}.
However, it is well known that inflation in SUGRA suffers from several serious problems.
For instance, when the scalar potential is dominated by F-term during inflation, the $\eta$-problem\cite{CSte} spoils the flatness of the inflaton potential and hence
a slow-roll condition is violated.\footnote{In Ref.\cite{Ste2} a proposal to cure this problem by quantum contribution is described.
There also exist proposals to cause successful inflation during F-term domination, by using particular Ans\"{a}tze for the K\"{a}hler potential\cite{KYY00,Pallis}.
See also the recent proposal\cite{KS10}, via embedding of $R^2$-inflation\cite{Sta} into supergravity, as a solution to the $\eta$-problem.}
If D-term dominates the scalar potential\cite{BD1996,LR} during inflation, one can escape from the above problem too (at least at the classical level).
However, the D-term inflation is known to very sensitive to the gauge charges, because inflaton lies in a vector superfield.

Though not directly related to above theories, the other well-known theory that may elegantly explain inflation exists;
it is Quantum Field Theory (QFT) in curved spacetime\cite{BDPT}.
Since inflation is expected to take place in very high-energy regime (just a few orders below the Planck scale), it seems to be natural to consider quantum effects of matter
onto the spacetime geometry.\footnote{The pioneering work of this subject by Starobinsky\cite{Sta} uses curved-space QFT.}
This theory may be thought of as the extension of the $f(R)$-theory of gravity\cite{SFFT}, which is the attractive alternative to GR.

An $f(R)$-gravity Lagrangian, regarded as a phenomenological proposal away from the original motivation (curved-space QFT),
is an arbitrary function of the Ricci scalar $R$, so it does not contain the derivative operator.
However, from the viewpoint of general covariance, it is allowed to include the derivative operator (d'Alembertian $\Box$, in particular) into the Lagrangian,
leading to the $f(R,\Box R)$-gravity that can be thought of as a natural generalization of the original $f(R)$-theory:
\be
f(R,\Box R) = R + \sum_{i=1}^{\infty}\alpha_{i}R^{i+1} + \sum_{j=1}^{\infty}\gamma_{j}R\Box^j R.\label{eqn:frgen}
\ee
There exists a \emph{physical} bonus of including the $\Box$-operator into the Lagrangian.
The derivative operator naturally induces a parallel transport, so the Ricci scalar can change from point to point when d'Alembertian $\Box$ is applied.
That is, our Lagrangian \eqref{eqn:frgen} may account for an inhomogeneity that may be present in a very early stage of the Universe.
Here we restrict ourselves to the \emph{linear} in the $\Box^j R$-term case: otherwise the theory would suffer from ghosts, as was pointed out in Ref.\cite{HOW}.
We will also restrict ourselves to the case of linear factor in the $\Box$-operator in the next section, for simplicity.
This type of Lagrangian \eqref{eqn:frgen} is often used in the context of inflationary and bouncing cosmologies\cite{GSS,A.etal,BMS,STT}.

The significance of the $R\Box R$-term on an inflationary scenario is re-examined in this paper, by adopting the notion of \emph{mutated} hybrid inflation in GR\cite{Ste3}.
In the next section we confirm that the theory described by the $f(R,\Box R)$-Lagrangian is equivalent to Einstein gravity (i.e. GR) coupled to two scalar fields\cite{Wands}.
We also show that a hybrid inflation can be realized in $f(R,\Box R)$-gravity by using a simple example.
In section \ref{sec:rescue}, we promote our toy model to a more realistic one, by demanding the stability of a waterfall field against a small perturbation.
Inflationary observables of our model are also estimated in that section.
In the final section we summarize the results and discuss some interesting issues related to our model.

\section{Embedding mutated hybrid inflation into $f(R,\Box R)$-model}\label{sec:Action}
In this paper we focus on the $f(R,\Box R)$-gravity theory with the action:
\be
I[g] = \frac{1}{2\gravcoup}\int d^{4}x \sqrt{-g}\;f(R,\Box R). \label{eqn:original-action}
\ee
The matter action is not considered in this paper.
Let us consider the following naive \emph{Ansatz}\cite{GSS,HOW}:
\be
f(R,\Box R) = R + \alpha R^2 + \gamma R\Box R, \label{eqn:fR1}
\ee
where $\alpha$ and $\gamma$ are some constants.
As is shown below, it is physically unacceptable Ansatz; it should be appended by another higher-order term in the scalar curvature.

As is well known, the action \eqref{eqn:original-action} with our choice \eqref{eqn:fR1} is dynamically equivalent to the scalar-tensor gravity
with two scalar fields (see, e.g. Ref\cite{Wands}),
\be
I[g,\Phi,\psi] = \frac{1}{2\gravcoup}\int d^{4}x \sqrt{-g}
                      \Bigl[ \Phi R + \gamma \psi\Box\psi - \left\{ \psi(\Phi - 1) - \alpha\psi^2 \right\} \Bigr]. \label{eqn:Jordan-action}
\ee
This equivalence is easily seen by considering a field equation of the $\Phi$-field.
Inserting $\psi = R$ that comes from $\delta S/\delta \Phi = 0$ into \eqref{eqn:Jordan-action}, we recover the original action with \eqref{eqn:fR1}.
According to the ordinary procedure, we carry out a frame change to the so-called Einstein frame given by
\nbe
g_{\mu\nu} \rightarrow \tilde{g}_{\mu\nu} = \Phi g_{\mu\nu}.
\nee
Then the action \eqref{eqn:Jordan-action} becomes
\begin{align}
I&[g,\phi,\psi] \notag\\
  &= \frac{1}{2\gravcoup}\int d^{4}x \sqrt{-g}
      \left[ R -\gravcoup(\pd\phi)^2 - \gamma e^{-\beta\kappa_4\phi}(\pd\psi)^2
                - e^{-2\beta\kappa_4\phi}\left\{ \psi\left(e^{\beta\kappa_4\phi} - 1\right) - \alpha\psi^2 \right\} \right], \label{eqn:Einstein-action}
\end{align}
where we have omitted the tilde for simplicity, and have reparametrized $\Phi$ as
\nbe
\Phi = e^{\beta\kappa_4\phi}; \quad \beta = \sqrt{\frac{2}{3}}.
\nee
Thus, in order for the theory to be ghost-free, a parameter $\gamma$ must satisfy
\be
  \gamma > 0.
\ee
As is desired, the action \eqref{eqn:Einstein-action} describes two scalar fields minimally coupled to Einstein gravity.
Note that the $\psi$-field, which we will identify as the waterfall field below, has a noncanonical kinetic term.
This is a generic feature of gravity theory derived from $f(R,\Box R)$-Lagrangian: only a single scalar acquires a canonical kinetic term with the physical sign.

Now we will focus on the potential term. It can be written as
\be
V(\phi,\psi) = \frac{1}{2\gravcoup}e^{-2\beta\kappa_4\phi}
                     \left[ -\alpha\left( \psi - \frac{e^{\beta\kappa_4\phi}-1}{2\alpha} \right)^2 + \frac{1}{4\alpha}\left(e^{\beta\kappa_4\phi}-1\right)^2 \right],
                     \label{eqn:pot}
\ee
where we assume
\be
\alpha > 0
\ee
via connection to the Starobinsky's model\cite{Sta}.\footnote{The $\alpha<0$ case is omitted from our discussion.}
Now we identify the $\phi$-field as the inflaton.
At each value of $\phi$, we assume that our waterfall field $\psi$ \emph{maximizes} the potential $V$, in order to generate the huge
potential energy that is required for inflation.
That is, the inflaton potential $V(\phi)$ is `propped up' by the $\psi$-field as
\be
\psi = \frac{e^{\beta\kappa_4\phi}-1}{2\alpha}.\label{eqn:waterfall}
\ee
Applying Eq.\eqref{eqn:waterfall} to the potential \eqref{eqn:pot}, we recover the Starobinsky's potential\cite{Sta}
\nbe
V(\phi) = \frac{1}{8\alpha\gravcoup}( 1 - e^{-\beta\kappa_4\phi} )^2,
\nee
but with a \emph{noncanonical} kinetic term $K(\phi)$, due to the contribution from the $\psi$-field,
\begin{align}
K(\phi) &= -\half(\pd\phi)^2 - \frac{\gamma e^{-\beta\kappa_4\phi}}{8\alpha^2\gravcoup}(\pd e^{\beta\kappa_4\phi})^2 \notag \\
           &= -\half\left( 1 + \frac{\gamma}{6\alpha^2}e^{\beta\kappa_4\phi} \right)(\pd\phi)^2.
           \label{eqn:kinterm}
\end{align}
That is, the inclusion of a $R\Box R$-term may be interpreted as a modification of the kinetic term of inflaton,
without altering the potential of $R^2$-inflation model.

\section{Inflationary observables of our model}\label{sec:rescue}
\begin{figure}
{\centering
\includegraphics[width=8.5cm]{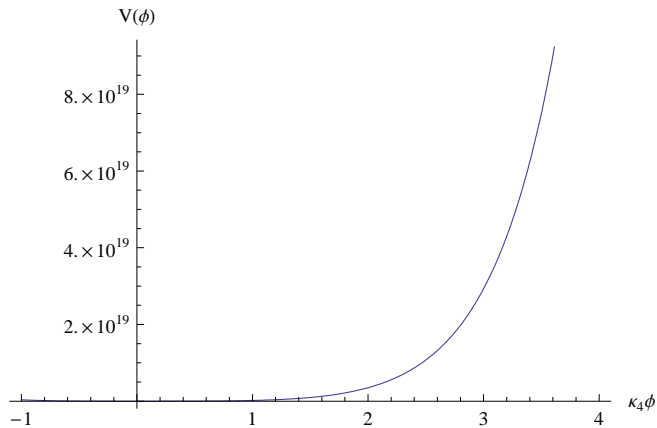}
\caption{An example of the inflaton potential $V(\phi)$.}
\label{fig:graph}}
\end{figure}

The configuration \eqref{eqn:waterfall} of a waterfall field corresponds to the apex of the potential $V$ in $\psi$-direction and is unstable against a small fluctuation.
Because $V$ is not bounded from below in $\psi$-direction, such a fluctuation would lead to a divergence $V \rightarrow -\infty$.
This implies that the Ansatz \eqref{eqn:fR1} is physically unacceptable.
So, we add the $R^4$-term to the Lagrangian \eqref{eqn:fR1}, whose role will be clarified later:
\be
I[g] = \frac{1}{2\gravcoup}\int d^{4}x \sqrt{-g}\Bigl( R + \alpha R^2 + \xi R^4 + \gamma R\Box R \Bigr),\label{eqn:modified-action}
\ee
which is dynamically equivalent to
\nbe
I[g,\Phi,\psi] = \frac{1}{2\gravcoup}\int d^{4}x \sqrt{-g}
                      \Bigl[ \Phi R + \gamma \psi\Box\psi - \left\{ \psi(\Phi - 1) - \alpha\psi^2 - \xi\psi^4 \right\} \Bigr].
\nee
Then the potential \eqref{eqn:pot} turns out to be
\be
V(\phi,\psi)
= \frac{1}{2\gravcoup}e^{-2\beta\kappa_4\phi}
   \left[ -\alpha\left( \psi - \frac{e^{\beta\kappa_4\phi}-1}{2\alpha} \right)^2 + \frac{1}{4\alpha}\left(e^{\beta\kappa_4\phi}-1\right)^2 - \xi\psi^4 \right],
   \label{eqn:mod-pot}
\ee
while the kinetic term does not change. Here we assume
\be
\xi < 0
\label{eqn:signcondition}
\ee
that gives the lower limit to the $\psi$-direction of the potential $V$. The problem stated above is thus resolved.
As before, we impose Eq.\eqref{eqn:waterfall} on a waterfall field. Then we have the inflaton potential (see Fig.~\ref{fig:graph})
\be
V(\phi) = \frac{1}{32\alpha^4\gravcoup}
              \left( 1 - e^{-\beta\kappa_4\phi} \right)^2\left[ 4\alpha^3 - \xi\left( e^{\beta\kappa_4\phi} - 1 \right)^2 \right].
\label{eqn:genpot}
\ee

Having established a physically reasonable potential \eqref{eqn:genpot}, our next task is to observe that our model gives reasonable inflation.
Well-known formalism for slow-roll inflation assumes scalar fields with canonical kinetic terms\cite{LL00}.
So we need to bring the kinetic term \eqref{eqn:kinterm} to the canonical one.
To do this, we use the techniques described below, instead of finding a exact solution
$\chi = \chi(\phi)$ s.t. $K(\phi) \rightarrow -\half(\pd\chi)^2$.\footnote{In fact, there \emph{exists} an exact solution that brings our kinetic term \eqref{eqn:kinterm}
to the canonical form. However, it is not very convenient for calculation of inflationary observables.}

In a small-$\phi$ region ($\kappa_4\phi \lesssim 1$), we may approximate our Lagrangian as
\begin{gather}
K(\phi) \simeq -\half(\pd\phi)^2, \notag \\
V(\phi) \simeq \frac{\beta^2}{8\alpha}\phi^2 - \frac{\beta^3\kappa_4}{8\alpha}\phi^3. \notag
\end{gather}
Thus the slow-roll conditions\cite{LL92}
\be
\varepsilon \ll 1, \quad
|\eta| \ll 1
\;\text{ where }\;
\varepsilon = \frac{1}{2\gravcoup}\left( \frac{V'}{V} \right)^2, \quad
\eta = \kappa_4^{-2}\frac{V''}{V},
\label{eqn:infcondition}
\ee
immediately give $\kappa_4\phi > 1$, and we have the contradiction with our initial assumption $\kappa_4\phi \lesssim 1$.
Hence, \emph{inflation cannot occur in this region.}
This fact allows us to focus only on a large-$\phi$ region, i.e. $\kappa_4\phi \gtrsim 1$.

In a large-$\phi$ region ($\kappa_4\phi \gtrsim 1$), the kinetic and potential terms are approximately given by
\begin{gather}
K \simeq -\half \cdot \frac{\gamma}{6\alpha^2}e^{\beta\kappa_4\phi}(\pd\phi)^2
   = -\half(\pd\varphi)^2,
   \notag \\
V \simeq \frac{-\xi}{32\alpha^4\gravcoup}\left( e^{2\beta\kappa_4\phi} - 4e^{\beta\kappa_4\phi} \right)
   = \frac{-\xi}{32\alpha^2\gamma}\left( \frac{(\alpha\kappa_4)^2}{\gamma}\varphi^4 - 4\varphi^2 \right)
   \label{eqn:large-potential}
\end{gather}
respectively, where we have defined a \emph{rescaled} inflaton
\be
\varphi = \frac{\gamma^{1/2}}{\alpha\kappa_4}e^{\beta\kappa_4\phi/2}.
\label{eqn:resinf}
\ee
The potential $V(\varphi)$ in terms of the rescalced inflaton $\varphi$ is similar to the Linde's chaotic one \cite{Linde83} of quartic order.
The difference is that we now have the potential that is \emph{polynomial} in $\varphi$.
Slow-roll conditions \eqref{eqn:infcondition} give us
\be
\kappa_4\varphi \gtrsim 2\sqrt{2 + \frac{\gamma}{\alpha^2}}.
\ee
The value $\varphi_\mathrm{end}$, where inflation ends, is also found from Eq.\eqref{eqn:infcondition}.
Solving $\varepsilon(\varphi_\mathrm{end}) = 1$, where there is a violation of slow-roll condition, gives
\be
\kappa_4\varphi_\mathrm{end} = 2\sqrt{2 + \frac{\gamma}{\alpha^2}}.
\ee
The scalar spectral index $n_\mathrm{s}$ in terms of slow-roll parameters\cite{LL92} is:
\begin{align}
n_\mathrm{s} &= 1 - 6\varepsilon + 2\eta \notag \\
                    &= 1 - \frac{24}{(\kappa_4\varphi)^2} - \frac{112(\gamma/\alpha^2)}{(\kappa_4\varphi)^4} + O\left((\kappa_4\varphi)^{-6}\right),
\label{eqn:index}
\end{align}
where the right hand side is evaluated at the horizon crossing of a physically interesting scale, as usual.
The number of $e$-foldings $N$ is
\begin{align}
N &= \gravcoup\int_{\varphi_\mathrm{end}}^{\varphi}\frac{V}{V'}d\varphi \notag \\
   &\simeq \frac{(\kappa_4\varphi)^2}{8} - \frac{2 + \gamma/\alpha^2}{2} - \frac{\gamma/\alpha^2}{2}\ln\varphi.
\end{align}
This equation can be solved \emph{approximately} as follows:
\be
(\kappa_4\varphi)^{-2} \cong \frac{1}{8N}\left( 1 - \frac{2+\gamma/\alpha^2}{2N} + \frac{(2+\gamma/\alpha^2)^2}{4N^2} \right) + O(N^{-4}).
\label{eqn:fieldvalue}
\ee
Inserting this equation into Eq.\eqref{eqn:index}, we find that the scalar spectral index of our model is given by
\be
n_\mathrm{s} \cong
1 - \frac{3}{N} + \frac{12-\gamma/\alpha^2}{4N^2} - \frac{(2 + \gamma/\alpha^2)(3 - 2\gamma/\alpha^2)}{2N^3}.
\label{eqn:ourindex}
\footnote{Note that the original Starobinsky's model\cite{Sta} predicts (see Refs.\cite{MC81,KKW10})
\nbe
n_\mathrm{s} \simeq 1- \frac{2}{N} + \text{(sub-leading terms)}.
\nee
The big difference from our model is the second term in the R.H.S. of Eq.\eqref{eqn:ourindex}. This is because our inflationary model is mainly dominated by $\varphi^4$-term.}
\ee
From this equation, we obtain the following condition that relates the two parameters:
\be
\gamma/\alpha^2 = \frac{1}{8}\left( -2 + N + \sqrt{196(1-N)+193N^2-64(1-n_\mathrm{s})N^3} \right).
\label{eqn:precondition}
\ee
An equation \eqref{eqn:precondition} itself gives an interesting speculation on the number of $e$-folds $N$.
For example, let us substitute the most recent WMAP7 observation for the scalar spectral index, $n_\mathrm{s} = 0.963 \pm 0.012$ (with $95\%$ CL)\cite{Komatsu.etal},
into this equation.
Then, the reality condition for a coefficient  $\gamma/\alpha^2$ implies that the $e$-folds $N$ of our model is bounded from above,
\begin{gather}
N < 61 \;(\text{when } n_\mathrm{s}=0.951), \notag\\
N < 81 \;(\text{when } n_\mathrm{s}=0.963), \notag\\
N < 120 \;(\text{when } n_\mathrm{s}=0.975).
\label{eqn:conditione-folds}
\end{gather}
This is a bonus of our model: without knowing the detail of reheating after inflation\cite{Reheating}, the upper limit on the $e$-folds $N$
(which is very sensitive to the $n_\mathrm{s}$) is found.

\begin{figure}
{\centering
\includegraphics[width=8.5cm]{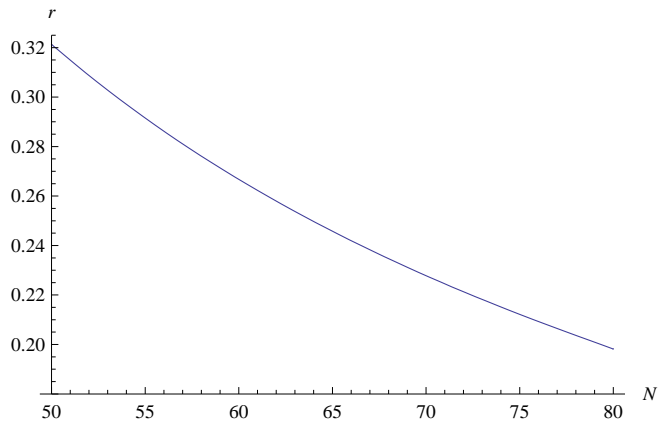}
\caption{A relation between the $e$-folds $N$ and the tensor-to-scalar ratio $r$, Eq.\eqref{eqn:ratio}, is depicted.}
\label{fig:ratio}}
\end{figure}

For concreteness, from now on we work with a central value that is indicated by WMAP7 observations, $n_\mathrm{s} \cong 0.963$.
Then the tensor-to-scalar ratio $r=16\varepsilon$\cite{LL92} is given by (see Fig.~\ref{fig:ratio})
\begin{align}
r &\cong \frac{16}{N}\left( 1 - \frac{1}{N} + \frac{2 + \gamma/\alpha^2}{N^2} \right)
\notag \\
   &= \frac{16}{N}\left\{ 1 - \frac{7}{8N} + \frac{1}{8N^2}\left( 14 + \sqrt{196(1-N)+193N^2-2.37N^3} \right) \right\}.
\label{eqn:ratio}
\end{align}
The WMAP7 observations (in combinations with BAO $+$ SN) of the tensor-to-scalar ratio, $r<0.20$\cite{Komatsu.etal}, severely constrain our model.
As Eq.\eqref{eqn:ratio} indicates, in order to be consistent with observations, the $e$-folds of our model should take $N\cong80$,
\emph{which is all but the maximum value of the} $N$ [\emph{see Eq.\eqref{eqn:conditione-folds}}].

Hence, we work with $N=80$, giving $r=0.198$.
Then Eq.\eqref{eqn:precondition} becomes
\be
\gamma/\alpha^2 = 20.4.
\label{eqn:condition1}
\ee
The COBE normalization $V^{1/4} = 0.027\varepsilon^{1/4}\kappa_4^{-1}$ at $k \cong k_\mathrm{pivot} = 0.002\mathrm{Mpc}^{-1}$\cite{Lyth97} gives another relation.
Applying a slow-roll parameter $\varepsilon$ of our model to this normalization, we find
\begin{align}
V &= \frac{5.3\times 10^{-7}}{N\kappa_4^4}
        \left\{ 1 - \frac{7}{8N} + \frac{1}{8N^2}\left( 14 + \sqrt{196(1-N)+193N^2-2.37N^3} \right) \right\} \notag\\
   &= \frac{6.6\times 10^{-9}}{\kappa_4^{4}},
   \label{eqn:pivot1}
\end{align}
where $N=80$ is used in the last line.
In the meanwhile, applying Eq.\eqref{eqn:fieldvalue} [combined with $N=80$ and Eq.\eqref{eqn:condition1}] to the potential \eqref{eqn:large-potential}, then it becomes
\be
V = \frac{35.3(-\xi)}{\gravcoup\alpha^4}.
\label{eqn:pivot2}
\ee
Combining Eqs.\eqref{eqn:pivot1} and \eqref{eqn:pivot2}, we obtain the other condition relating two parameters:
\be
\frac{-\xi}{\alpha^4} = \frac{1.9 \times 10^{-10}}{\gravcoup}.
\label{eqn:condition2}
\ee
Note that we are left with only a \emph{single} free parameter [with the one sign; see, e.g. Eq.\eqref{eqn:signcondition}].

\section{Conclusions and discussions}\label{sec:conc}
We studied classical dynamics of inflation derived from $f(R,\Box R)$-gravity.
We observed that multiscalar nature of this gravity theory allowed us to apply it to hybrid-type inflationary model-building.
Under the assumption that a waterfall field supports the inflaton potential, we found that our model exhibits a large-field inflationary behavior,
and is similar to Linde's chaotic model of the quartic order\cite{Linde83}.
By using the observational data of WMAP satellite\cite{Komatsu.etal} and the COBE normalization \cite{Lyth97}
we found two conditions, Eqs.\eqref{eqn:condition1} and \eqref{eqn:condition2}, that should be satisfied by the coefficients in the action.
They severely restrict the parameter space of our inflationary model.
Notable point is that we are left with only a \emph{single} free parameter with the odd sign.

It is also found that the limit on the gravitational wave production $r<0.20$\cite{Komatsu.etal} together with Eq.\eqref{eqn:conditione-folds} \emph{almost fixes}
the number of $e$-folds $N$ of our model to be eighty (when a central value for the scalar spectral index $n_\mathrm{s}=0.963$ is used).
This is essentially due to the fact that our potential, \emph{under a large-field approximation}, is dominated by
the \emph{quartic}-order term in the rescaled inflaton \eqref{eqn:resinf}, which is widely known as a `creator' of large gravitational waves.
This fact may indicate that we should treat our original potential \eqref{eqn:genpot}
without any approximation, and construct a more viable model that creates smaller gravitational waves.
Exact analysis of our potential may improve the situations.
It may also happen that the inclusion of only the linear factor in the $\Box$-operator is not sufficient
to build a realistic model of inflation and we may have to include more higher-order terms in d'Alembertian.

The study of a cosmological perturbation theory\cite{KS84} is of great importance.
The PLANCK satellite mission\cite{pla} is expected to remove the ambiguity in the primordial non-Gaussianity\cite{Mal03} of cosmic perturbations.
The WMAP gives two distinct conditions on the non-Gaussianity parameter $f_\mathrm{NL}$\cite{Komatsu.etal}.
The perturbation theory of our model is expected to give certain restrictions on the parameter space, so that it will be possible to test our model soon.

\vskip 10pt

{\centering{\subsection*{Acknowledgements}}}
The author is very grateful to A. A. Starobinsky and S. V. Ketov for stimulating discussions, valuable comments and references.


\vskip 30pt




\begin{thebibliography}{99}
\bibitem{Guth}
  A. H. Guth,
  Phys. Rev. \textbf{D23}, 347 (1981).
\bibitem{Komatsu.etal}
  E. Komatsu \emph{et al.},
  arXiv:1001.4538 [astro-ph]
\bibitem{Erd}
  J. Erdmenger (Ed.),
  \emph{String cosmology: Modern string theory concepts from the Big Bang to cosmic structure},
  Wiley-VCH (2009).
\bibitem{MSYY94}
  H. Murayama, H. Suzuki, T. Yanagida and J. Yokoyama,
  Phys. Rev. \textbf{D50}, 2356 (1994); arXiv:hep-ph/9311326
\bibitem{KL10}
  R. Kallosh and A. Linde,
  J. Cosmol. Astropart. Phys. 11 (2010) 011; arXiv:1008.3375 [hep-th] 
\bibitem{CSte}
  In fact this kind of problem arises \emph{any} real scalar field in SUGRA during inflation.
  The importance of this problem to inflationary model building is pointed out in, e.g. \\
  E. J. Copeland \emph{et al.},
  Phys. Rev. \textbf{D49}, 6410 (1994);
  arXiv:astro-ph/9401011; \\
  E. D. Stewart,
  Phys. Rev. \textbf{D51}, 6847 (1995);
  arXiv:hep-ph/9405389
\bibitem{Ste2}
  E. D. Stewart, \\
  Phys. Lett. \textbf{B391}, 34 (1997);
  arXiv:hep-ph/9606241; \\
  Phys. Rev. \textbf{D56}, 2019 (1997);
  arXiv:hep-ph/9703232
\bibitem{KYY00}
  M. Kawasaki, M. Yamaguchi and T. Yanagida,
  Phys. Rev. \textbf{D63}, 103514 (2001); arXiv:hep-ph/0011104
\bibitem{Pallis}
  C. Pallis,
  J. Cosmol. Astropart. Phys. 04 (2009) 024;
  arXiv:0902.0334 [hep-ph]
\bibitem{KS10}
  S. V. Ketov and A. A. Starobinsky,
  arXiv:1011.0240 [hep-th]
\bibitem{BD1996}
  P. Binetruy and G. R. Dvali,
  Phys. Lett. \textbf{B388}, 241 (1996);
  arXiv:hep-ph/9606342
\bibitem{LR}
  D. H. Lyth and A. Riotto,
  Phys. Lett. \textbf{B412}, 28 (1997);
  arXiv:hep-ph/9707273
\bibitem{SFFT}
  For example, see following recent reviews: \\
  T. P. Sotiriou and V. Faraoni,
  Rev. Mod. Phys. \textbf{82}, 451 (2010);
  arXiv:0805.1726 [gr-qc]; \\ 
  A. De Felice and S. Tsujikawa,
  Living Rev. Rel. \textbf{13}, 3 (2010);
  arXiv:1002.4928 [gr-qc]; \\
  S. Nojiri and S. D. Odintsov,
  arXiv:1011.0544 [gr-qc]
\bibitem{Sta}
  A. A. Starobinsky,
  Phys. Lett. \textbf{B91}, 99 (1980).
\bibitem{BDPT}
  N. D. Birrell and P. C. W. Davies,
  \emph{Quantum fields in curved space},
  Cambridge University Press (1982); \\
  L. Parker and David Toms,
  \emph{Quantum Field Theory in Curved Spacetime: Quantized Fields and Gravity},
  Cambridge University Press (2009).
\bibitem{HOW}
  A. Hindawi, B. A. Ovrut and D. Waldram,
  Phys. Rev. \textbf{D53}, 5597 (1996);
  arXiv:hep-th/9509147
\bibitem{GSS}
  S. Gottlober, H. J. Schmidt and A. A. Starobinsky,
  Class. and Quant. Grav. \textbf{7}, 893 (1990)
\bibitem{A.etal}
  L. Amendola \emph{et al.},
  Class. and Quant. Grav. \textbf{10}, L43 (1993).
\bibitem{BMS}
  T. Biswas, A. Mazumdar and W. Siegel,
  J. Cosmol, Astropart. Phys. 03 (2006) 009;
  arXiv:hep-th/0508194
\bibitem{STT}
  M. Skugoreva, A. Toporensky and P. Tretyakov,
  arXiv:1007.3365 [gr-qc]
\bibitem{Ste3}
  E. D. Stewart,
  Phys. Lett. \textbf{B345}, 414 (1995);
  arXiv:astro-ph/9407040
\bibitem{Wands}
  D. Wands,
  Class. and Quant. Grav. \textbf{11}, 269 (1994);
  arXiv:gr-qc/9307034
\bibitem{LL00}
  A. R. Liddle and D. H. Lyth,
  \emph{Cosmological Inflation and Large-Scale Structure}
  (Cambridge University Press, 2000).
\bibitem{LL92}
  A. R. Liddle and D. H. Lyth,
  Phys. Lett. \textbf{B291}, 391 (1992);
  arXiv:astro-ph/9208007
\bibitem{Linde83}
  A. Linde,
  Phys. Lett. \textbf{B129}, 177 (1983).
\bibitem{MC81}
  V. F. Mukhanov and G. V. Chibisov, JETP Lett. \textbf{33}, 532 (1981).
\bibitem{KKW10}
  S. Kaneda, S. V. Ketov and N. Watanabe,
  Mod. Phys. Lett. \textbf{A25}, 2753 (2010);
  arXiv:1001.5118 [hep-th]
\bibitem{Reheating}
  A. Albrecht, P. J. Steinhardt, M. S. Turner and F. Wilczek,
  Phys. Rev. Lett. \textbf{48}, 1437 (1982); \\
  L. Kofman, A. D. Linde and A. A. Starobinsky,
  Phys. Rev. Lett. \textbf{73}, 3195 (1994);
  arXiv:hep-th/9405187
\bibitem{Lyth97}
  D. H. Lyth,
  Phys. Rev. Lett. \textbf{78}, 1861 (1997);
  arXiv:hep-ph/9606387
\bibitem{KS84}
  H. Kodama and M. Sasaki,
  Prog. Theor. Phys. Suppl. \textbf{78}, 1 (1984).
\bibitem{pla}
  http://www.rssd.esa.int/index.php?project=Planck
\bibitem{Mal03}
  J. M. Maldacena,
  J. High Energy Phys. 05 (2003) 013;
  arXiv:astro-ph/0210603
\end{thebibliography}
\end{document}